\documentclass[sigconf]{acmart}
\usepackage{titlesec}
\AtBeginDocument{%
  }

\setcopyright{none}

\acmConference[CHI'25 Workshop on Tools for Thought]{Tools for Thought: Research and Design for Understanding, Protecting, and Augmenting Human Cognition with Generative AI on CHI 2025 Workshop}{April 26,2025}{Yokohama, JAPAN}
  
\renewcommand\footnotetextcopyrightpermission[1]{}

\acmISBN{978-1-4503-XXXX-X/2018/06}

\usepackage{booktabs}
\usepackage{multirow}
\usepackage{array}
\usepackage{geometry}

\begin{document}

\title[Beyond Tools]{Beyond Tools: Understanding How Heavy Users Integrate LLMs into Everyday Tasks and Decision-Making}

\author{Eunhye Kim}
\email{gracekim027@snu.ac.kr}
\affiliation{%
  \institution{Seoul National University}
  \city{Seoul}
  \country{Republic of Korea}
}

\author{Kiroong Choe}
\email{krchoe@hcil.snu.ac.kr}
\affiliation{%
  \institution{Seoul National University}
  \city{Seoul}
  \country{Republic of Korea}
}

\author{Minju Yoo}
\email{minjuu613@ewhain.net}
\affiliation{%
  \institution{Ewha Womans University}
  \city{Seoul}
  \country{Republic of Korea}
}

\author{Sadat Shams Chowdhury}
\email{sadatshams@kaist.ac.kr}
\affiliation{%
  \institution{School of Computing, KAIST}
  \city{Daejeon}
  \country{Republic of Korea}
}

\author{Jinwook Seo}
\email{jseo@hcil.snu.ac.kr}
\affiliation{%
  \institution{Seoul National University}
  \city{Seoul}
  \country{Republic of Korea}
}

\renewcommand{\shortauthors}{Kim et al.}

\begin{abstract}
  Large language models (LLMs) are increasingly used for both everyday and specialized tasks. While HCI research focuses on domain-specific applications, little is known about how heavy users integrate LLMs into everyday decision-making. Through qualitative interviews with heavy LLM users (n=7) who employ these systems for both intuitive and analytical thinking tasks, our findings show that participants use LLMs for social validation, self-regulation, and interpersonal guidance, seeking to build self-confidence and optimize cognitive resources. These users viewed LLMs either as rational, consistent entities or average human decision-makers. Our findings suggest that heavy LLM users develop nuanced interaction patterns beyond simple delegation, highlighting the need to reconsider how we study LLM integration in decision-making processes.

\end{abstract}



\maketitle
\section{Introduction}
The ubiquitous question "How did I do this before ChatGPT?" has become a cultural touch point, highlighting how Large Language Models (LLMs) have gradually permeated people's everyday lives. While initially introduced as general-purpose chatbots, LLMs have been adopted in unexpectedly diverse ways \cite{chkirbene2024applications}. These systems now play multiple roles in decision-making processes and tasks, ranging from information providers to triggers for human self-reflection \cite{kim2022bridging, kmmer2024effects}. This widespread integration has raised the question about how users develop dependencies on and relationships with these AI systems \cite{he2025conversational}.

Previous Human-Computer Interaction (HCI) research has extensively examined domain-specific LLM applications \cite{jin2024teach, liu2024selenite}. These studies have yielded insights into specialized use cases and led to targeted interaction design improvements. However, the broader impact of LLMs on everyday decision-making and tasks remains under-explored. As users increasingly integrate these tools into their daily routines, understanding the tangible impacts of habitual use becomes crucial \cite{kmmer2024effects}. 

Recent studies have attempted to measure the impact of LLM use through quantitative metrics such as task performance and decision accuracy \cite{kim2025fostering}. However, these measurement-based approaches cannot fully capture how people delegate everyday decisions to LLMs or the resulting meta-cognitive effects. Furthermore, everyday decisions encompass a broad spectrum of choices, from routine task management to social interaction planning \cite{eigner2024determinants, dhami2012cct}, making them challenging to examine through purely quantitative and task-specific approaches.

To address this gap, we conducted a qualitative study examining heavy LLM users who regularly rely on these systems for everyday decisions and tasks. Through interviews and analysis, we explored how these users integrate LLMs into their decision-making processes, what types of decisions they choose to delegate, and how this delegation affects their cognitive patterns and decision-making confidence. Our study addressed three research questions: 

\begin{itemize}
\item RQ1: How do heavy LLM users integrate LLMs into their everyday decision-making process?
\item RQ2: What underlying needs do heavy LLM users seek to fulfill through LLM assistance?
\item RQ3: How do heavy LLM users conceptualize and evaluate their relationship with LLMs?
\end{itemize}

Through these research questions, we examine three key aspects of heavy LLM use. RQ1 explores emergent use cases and notable patterns in how users incorporate LLMs into their decision-making processes. RQ2 investigates the fundamental motivations and needs that drive sustained LLM usage. RQ3 examines how users develop their mental models of LLMs and reflect on their extensive interaction with these systems.
\section{Methods}
\subsection{Data Gathering}
We identified heavy LLM users through a screening survey followed by semi-structured interviews. Heavy users were defined as individuals who regularly employed LLMs beyond work tasks, incorporating them into everyday decision-making processes and delegating various aspects of their decision-making to these systems.

\subsubsection{Participant Recruitment}
To identify heavy LLM users who regularly use these systems for everyday decisions, we sought participants who delegated decision-making tasks across both System 1 and System 2 thinking, as well as the spectrum between these two systems. To ensure comprehensive coverage of decision-making types, we based our screening survey on Hammond's Cognitive Continuum Theory (CCT) \cite{hammond2000human}. CCT classifies decision-making into four modes based on the balance between System 1 and System 2 thinking: Pure Intuition, Aided Intuition, Quasi-Rational Intuition, and Quasi-Rational Analysis. We developed 5-6 representative tasks for each mode. Examples include "deciding whether to speak up in class or meetings" (Pure Intuition), "choosing a recipe to cook" (Aided Intuition), "addressing interpersonal conflicts" (Quasi-Rational Intuition), "selecting elective courses" (Quasi-Rational Intuition), and "determining project scope" (Quasi-Rational Analysis). Survey participants indicated whether they had used LLMs for each task with three response options: "Yes, I have done this task with LLMs," "No, but I have made decisions like this before," or "Not Applicable to me." The complete task set for each mode is provided in Appendix~\ref{appendix:survey-tasks}.

\begin{table*}[htbp]
\begin{tabular}{llllcl}
\toprule
\textbf{Number} & \textbf{Age} & \textbf{Gender} & \textbf{Vocation} & \textbf{Tasks Done with LLMs} & \textbf{Usage} \\
\midrule
P1  & 36 & Male & Teacher (10 years experience) & 14 & 3-7 times per week \\
P2  & 25 & Female & 2nd Year Undergraduate (Environmental Science) & 17 & 3-7 times per week \\
P3  & 39 & Male & PhD Student / Part-time Lecturer & 18 & Daily \\
P4  & 20 & Female & 2nd Year Undergraduate (Economics) & 14 & Daily \\
P5  & 24 & Female & Job Seeker (Recent Graduate) & 13 & Daily \\
P6  & 26 & Male & 2nd Year PhD Student (Mechanical Engineering) & 18 & Daily \\
P7  & 20 & Male & 2nd Year Undergraduate (Computer Science) & 14 & Daily \\
\midrule
\textbf{N=7} & \textbf{27.1} & \textbf{M/F: 4/3} & & \textbf{15.4} & \textbf{Daily/Weekly: 5/2} \\
\bottomrule
\end{tabular}
\caption{Participant Profiles and LLM Usage}
\label{tab:participant-data}
\end{table*}

\subsubsection{Participants}
From 78 survey responses, we identified heavy users using three criteria. Participants needed to indicate LLM use in at least 13 of the 22 tasks, show balanced LLM use across different decision-making modes, and demonstrate LLM use for both work-related and personal tasks (e.g., recipe selection or outfit choices). Seven participants met these criteria (Table~\ref{tab:participant-data}).

\subsection{Semi-Structured Interview and Analysis}
We conducted an hour-long semi-structured interviews with each participant. Sessions began with an overview of the study's goals and an acknowledgment that there were no predetermined correct answers. The interview protocol addressed three main themes: participants' reasons for using LLMs in specific scenarios from the survey, their general motivation and usage patterns for LLMs, and their perceived changes in decision-making processes since adopting LLMs. Participants received KRW 15,000 (approximately 11 USD) as compensation. For the analysis, two researchers individually coded the interview transcripts using open coding. Then, a team of three researchers discussed the outcomes to resolve discrepancies and generate, review, and iterate on themes.

\section{Results}
\subsection{RQ1: How do heavy LLM users integrate LLMs into their everyday decision-making practices?}
Our first research question explored how heavy users incorporate LLMs into their decision-making processes. Through screening forms and interviews, we discovered diverse applications ranging from quick validations to complex relationship advice. The most common use cases included validation for social appropriateness and context understanding, self-regulation in purchase decisions, and interpersonal guidance.

\subsubsection{Case 1: Social Appropriateness and Context Understanding}
Participants used LLMs to navigate social situations where they felt uncertain about appropriate behavior. This pattern spanned from immediate communication decisions to long-term social planning. Participants primarily sought to align their actions with social norms, especially when they lacked confidence in their social judgment.

Six participants (P2-P7) used LLMs to validate their responses during time-sensitive situations like answering professors' questions or contributing to meetings. Rather than using search engines, they quickly shared context, for instance by giving the class material itself, and proposed their responses with LLMs to assess appropriateness. As P6 explained, they wanted to ensure responses "made sense in the provided context." This validation either boosted their confidence to respond or helped them avoid potentially awkward or disruptive contributions. Participants particularly valued avoiding responses that might interrupt the meeting flow with contextually inappropriate or "wrong" comments.

Participants also sought LLM guidance for context-appropriate attire. P2, new to academia, consulted ChatGPT about conference dress codes, specifically seeking validation for her preference of wearing jeans. P3 used LLMs to help choose appropriate attire for meeting a former partner. In both scenarios, participants used LLMs to ensure their choices aligned with social expectations.

P3, who self-identified as having lower social awareness, regularly used LLMs to navigate everyday social situations. When selecting gifts, he found value in ChatGPT's specific follow-up questions about the recipient, such as "whether this person usually wears gold or silver." Similarly, when deciding between wine or champagne for a party, he appreciated how LLMs considered multiple factors including the event's purpose, intended mood, and the social implications of each beverage choice. He noted that these interactions helped him recognize the complex factors involved in social decisions.

\subsubsection{Case 2: Self-Regulation in Purchase Decisions}
Two participants (P2, P7) employed LLMs as a tool for self-regulation during impulse purchases, though with divergent expectations and outcomes. P2 used LLM feedback to curb impulsive buying, stating, "I knew buying this item was impulsive and irrational, and just needed to hear it out from another being." This led to her avoiding the purchase. P7, however, leveraged LLMs to justify his impulse purchases, noting that "LLMs such as ChatGPT have a tendency of saying what the audience wants to hear." He deliberately chose LLMs over friends for purchase validation, knowing they would help rationalize his impulse buying decisions.

\subsubsection{Case 3: Interpersonal Guidance through LLMs}
Five participants (P2, P3, P4, P6, P7) sought LLM assistance for interpersonal challenges, though their approaches and expectations varied. While P6 used LLMs primarily as an emotional outlet during a conflict, others engaged in deeper social problem-solving. P2 and P3 used LLMs for self-reflection, asking questions like "Am I being too sensitive?". P3 described a particularly striking experience where ChatGPT analyzed his past conflict with an advisor. While the LLM validated some of his concerns about the situation, it simultaneously challenged him by pointing out how his confrontational response to the professor conflicted with traditional conservative ethical principles in Korean academic contexts - a dual perspective that he found profoundly unexpected. P4 developed a particularly significant relationship with ChatGPT, using it as a confidant for relationship advice. She valued its ability to help her understand her boyfriend's perspective, interpret text messages, and craft responses. P4 specifically appreciated the balance between neutral advice and emotional support, noting that the LLM could maintain objectivity while remaining sympathetic to her situation. Participants preferred LLMs over human advisors for these personal matters, citing the absence of social pressure and judgment. They noted that LLMs offered comprehensive social guidance while considering individual context, without the self-consciousness associated with seeking human advice.

\subsection{RQ2: What underlying needs do heavy LLM users seek to fulfill through LLM assistance?}
Our analysis implied three fundamental needs driving users' LLM engagement: boosting decision-making confidence, validating choices, and improving cognitive efficiency by delegating challenging tasks.

\subsubsection{Gaining Self-Confidence through Validation}
Participants often consulted LLMs not to make decisions, but to gain confidence in choices they were already inclined to make. P2 and P4 sought confirmation for preliminary decisions across various situations, including outfit choices, housing options, and course schedules. Rather than seeking help with the initial decision-making process, they would present their analysis of options to the LLM, seeking reassurance to overcome their uncertainty. This pattern of seeking reassurance extended to impulse purchases, where P2 and P7 approached LLMs with specific expectations. As both noted, "I just wanted to hear those words from someone"---revealing how users specifically sought out either permission or restraint from LLMs to support their predetermined decisions.

\subsubsection{Finding the "Right" Answer}
Participants utilized LLMs to consider as many options and diverse sources of information to reach optimal decisions, and to check if their thoughts were the "right" and "optimal" answer. For instance, participants would validate their ideas before meetings or use LLMs to organize options for trip planning and gift buying. They believed LLMs would consider comprehensive contextual information to provide the "right" answer. This aligned with participants' views of ideal decision-making - to consider as many options as possible (P1, P4, P5, P6). Participants leveraged LLMs to strengthen this approach by having the models organize different options, though some (P5) expressed concern that their decision-making might be confined to LLM-suggested options, potentially missing other alternatives.

\subsubsection{Optimizing Cognitive Resources through Task Delegation}
Participants completely delegated tasks they saw as not worth their mental effort to LLMs, seeking to free themselves from such responsibilities while preserving their cognitive energy for more valuable activities. The definition of "lower priority" varied among users in notably subjective ways. P6, a graduate student, chose to delegate course assignments to LLMs while engaging deeply with research work, despite both being essential academic responsibilities. P7 outsourced travel planning but enthusiastically handled laptop selection personally, citing a personal interest in electronics despite both tasks being similar to consumer decisions. P3, another graduate student, delegated lab-related social interactions while actively engaging in research ideation, showing how even within the same professional context, participants made highly personal choices about which tasks to delegate. In these cases, participants accepted and implemented LLM suggestions without seeking additional information or engaging in further analysis.

P6, a PhD student with a demanding schedule who described himself as "lacking energy," noted that he had developed a task-ranking system to identify which activities could be fully delegated to LLMs, requiring minimal personal mental effort. He prioritized delegating administrative duties and course assignments to LLMs in order to preserve his cognitive resources for higher-priority research work. This strategy emphasized maintaining cognitive efficiency for important tasks, even if it meant accepting lower-quality outcomes for less critical activities. Similarly, P3 applied this approach to social decisions, seeking satisfactory results while minimizing mental effort. As P6 explained, this deliberate trade-off between cognitive efficiency and output quality was specifically designed for lower-priority tasks.

\subsection{RQ3: How do heavy LLM users conceptualize and evaluate their relationship with LLMs?}
We examined users' mental models of LLMs and how these perceptions influenced their decision-making delegation patterns. Our analysis explored both their current usage patterns and their anticipated future reliance on LLMs for decision support.

\subsubsection{Mental Models of LLMs}
\paragraph{LLMs as Rational and Consistent Beings}
Participants predominantly conceptualized LLMs as consistent and decisive entities. P2 exemplified this by delegating complex decisions, such as course selection and housing choices, to LLMs. Unlike cases where participants used LLMs to compare options, P2 delegated entire decisions to LLMs when she struggled with decisiveness or questioned the optimality of her preferences. She valued ChatGPT's "perseverance" and rational nature, contrasting it with humans who might give inconsistent advice based on timing or previous interactions. Similarly, P4, who integrated LLMs into various aspects of her life, appreciated their dual capability of providing both rational guidance and perspective-aligned advice. She particularly valued that ChatGPT could offer both objective analysis and personally tailored guidance.

\paragraph{LLMs as Average Decision-Makers}
A contrasting mental model emerged where participants (P3, P7) viewed LLMs as "average and general decision-makers." P3, who felt he had below-average social understanding, relied on LLMs for social decisions, believing their training on vast data would provide at least average human-level responses. P7 valued this "averageness" for its guaranteed minimal competence with minimal time investment. However, this perceived averageness became a limitation for tasks requiring creativity and depth, such as research ideation or CV writing (P3, P5, P6, P7). Participants described LLM outputs as "exemplary," "general," and "uniform," leading them to seek alternative approaches for creative tasks. Some turned to colleagues (P3), others avoided LLM influence (P6), while P7 developed a strategy of asking LLMs to pose questions before starting tasks, optimizing their capabilities within recognized limitations.

\subsubsection{Self-Reflection on LLM Usage}
Participants demonstrated awareness of their extensive LLM use, often viewing it critically upon reflection during interviews. Their concerns spanned both societal and personal dimensions. On a personal level, some participants (P5, P6) worried about a potential decline in their creative and problem-solving abilities. P2 described consciously setting limits on which suggestions to follow. She noted that while ChatGPT provided the decisions, she ultimately bore the responsibility, expressing that it felt like "her boundaries were being invaded" and emphasizing that "I should be responsible." P3 expressed particular concern about the potential erosion of human agency, especially for ethical issues: "To maintain our humanity and human nature, we need to constantly verify and struggle with our own humanness. But because I find it tiresome and troublesome, I end up letting machines do it for me. If this happens on a real social scale, it's no joke." 

Despite these concerns, when asked about potentially reducing their LLM usage, participants indicated plans to maintain or expand their use, including diversifying usage across different domains. As P6 explained, while being worried about diminishing problem-solving skills, he also noted, "The problem-solving skills of LLMs will develop at a much faster speed than my own, so it would be better to try to figure out how to use LLMs better, which is another form of problem-solving." P7 also expressed a more optimistic view, stating, "Unconditionally accepting everything the LLM says isn't good, and it won't help us get better at making decisions. That's why I try to find different ways to use LLMs, like asking them to ask me questions that make me think more deeply. I think this kind of approach shows how we can use LLMs better in the future."
\section{Discussion}
Our study reveals how heavy users integrate LLMs into their daily tasks through distinct patterns. Rather than simple tool usage, participants demonstrated sophisticated cognitive offloading strategies that transformed their decision-making processes. In our study, we observed participants delegating social and interpersonal reasoning to LLMs, suggesting ways users might leverage AI collaboration to support their social cognition processes.

Participants' mental models of LLMs directly influenced their cognitive strategies---those viewing LLMs as rational entities engaged in cognitive complementarity by leveraging LLM capabilities where they perceived personal limitations, while those viewing LLMs as average decision-makers used cognitive benchmarking, establishing baseline standards while reserving higher-order tasks for themselves.

This raises questions for future research on redefining how we conceptualize and measure over-reliance on LLMs. Current metrics typically assess over-reliance through simplified quantitative measures in controlled settings, primarily focusing on users' acceptance rates of LLM outputs ~\cite{bo2024rely, kim2024rely}. However, our findings reveal more complex patterns of engagement. Participants did not blindly adopt LLM outputs, even in cases where they eventually accepted them. Instead, participants demonstrated thoughtful delegation strategies, using LLMs to validate existing decisions, automate routine tasks, or navigate unfamiliar situations. The critical concern was not users' acceptance of LLM outputs, but rather instances where users adopted LLM reasoning without exploring alternative perspectives. Future research should expand the definition of over-reliance beyond simple acceptance rates to examine how users critically engage with alternative lines of reasoning.

Another key direction for future research involves capturing diverse user contexts. Our participants valued the ability of LLMs to extract necessary contextual information when not initially provided. They appreciated that they could receive meaningful responses without extensively explaining background information, even for context-heavy topics like relationship advice. Future research should explore ways to incorporate multi-modal inputs beyond text-based interactions, allowing users to convey context through various channels. Additionally, LLMs' ability to elicit implicit user intentions without explicit prompting is crucial, as demonstrated by recent advances in reasoning-focused LLM architectures that can proactively identify and address underlying user needs.

The development of active usage patterns with LLMs appeared more prominent among younger users who had less experience managing tasks without these systems. Participants with extensive pre-LLM experience maintained clearer boundaries and showed greater awareness of system limitations. In contrast, users with less experience with LLMs demonstrated fewer reservations, viewing LLM interaction itself as a skill and actively developing their prompting strategies. Conducting design studies focused on younger generations, to better understand and support these emerging interaction patterns represents a crucial direction for future research.

\bibliographystyle{ACM-Reference-Format}
\bibliography{references}

\appendix
\section{Screening Survey Tasks}\label{appendix:survey-tasks}
The following tasks were presented with three response options:
\begin{itemize}
    \item Yes, I have used LLMs for this task
    \item No, but I have made decisions like this before
    \item Not applicable to me
\end{itemize}

\subsection{Mode 1: Pure Intuition Tasks}
\begin{enumerate}
    \item Deciding whether to speak up during class or meetings
    \item Responding quickly to text messages
    \item Making instant comments in Zoom chat
    \item Responding promptly to urgent emails
    \item Responding immediately to questions like "Does anyone know this?" in group chats
\end{enumerate}

\subsection{Mode 2: Aided Intuition Tasks}
\begin{enumerate}
    \item Composing appropriate email content for professors or colleagues/supervisors considering the current situation
    \item Deciding what content to include in a 5-minute group presentation
    \item Deciding what to cook with ingredients available in the refrigerator
    \item Planning how to structure a one-page assignment
    \item Planning outfit combinations with clothes in the wardrobe
\end{enumerate}

\subsection{Mode 3: Quasi-Rational Intuition Tasks}
\begin{enumerate}
    \item Composing responses to sensitive work emails
    \item Deciding whether to address issues with someone you frequently encounter
    \item Deciding whether to purchase items on sale
    \item Deciding which elective courses to take
    \item Deciding whether to participate in optional workshops
    \item Planning initial project scope
\end{enumerate}

\subsection{Mode 4: Quasi-Rational Analysis Tasks}
\begin{enumerate}
    \item Selecting courses for the next semester
    \item Deciding on housing/dormitory for future residence
    \item Planning travel itineraries
    \item Selecting laptops/tablets for school or work
    \item Making career-related decisions such as graduate school applications
    \item Planning in-depth projects or research papers
\end{enumerate} 
\end{document}